\documentclass[aip,preprint,superscriptaddress,floatfix]{revtex4-1}

\usepackage{mathptmx}
\usepackage[scaled=0.9]{helvet}
\usepackage[utf8]{inputenc}
\usepackage{graphicx}
\usepackage[dvipsnames]{xcolor}
\hypersetup{colorlinks,citecolor=blue}

\usepackage{textcomp}
\usepackage{ifsym}
\usepackage{mathrsfs}
\usepackage{sidecap}
\usepackage[
,textwidth=17.5cm
,textheight=23.5cm
,verbose
,dvips
]{geometry}
\usepackage{xspace}

\newcommand{\celsius}{$^{\circ}$C\xspace}

\begin{document}

\title{N-polar GaN/AlN resonant tunneling diodes}
\author{YongJin~Cho}
\thanks{These authors contributed equally to this work.}
\email[Author to whom correspondence should be addressed: Electronic mail: ]{yongjin.cho@cornell.edu}
\affiliation{School of Electrical and Computer Engineering, Cornell University, Ithaca, New York 14853, USA}
\author{Jimy~Encomendero}
\thanks{These authors contributed equally to this work.}
\email[Electronic mail: ]{jje64@cornell.edu}
\affiliation{School of Electrical and Computer Engineering, Cornell University, Ithaca, New York 14853, USA}
\author{Shao-Ting~Ho}
\affiliation{Department of Materials Science and Engineering, Cornell University, Ithaca, New York 14853, USA}
\author{Huili~Grace~Xing}
\affiliation{School of Electrical and Computer Engineering, Cornell University, Ithaca, New York 14853, USA}
\affiliation{Department of Materials Science and Engineering, Cornell University, Ithaca, New York 14853, USA}
\affiliation{Kavli Institute for Nanoscale Science, Cornell University, Ithaca, New York 14853, USA}
\author{Debdeep~Jena}
\affiliation{School of Electrical and Computer Engineering, Cornell University, Ithaca, New York 14853, USA}
\affiliation{Department of Materials Science and Engineering, Cornell University, Ithaca,
New York 14853, USA}
\affiliation{Kavli Institute for Nanoscale Science, Cornell University, Ithaca, New York 14853, USA}

\begin{abstract}
N-polar GaN/AlN resonant tunneling diodes are realized on single-crystal N-polar GaN bulk substrate by plasma-assisted molecular beam epitaxy growth. The room-temperature current-voltage characteristics reveal a negative differential conductance (NDC) region with a peak tunneling current of $6.8\pm 0.8$~kA/cm$^2$ at a forward bias of $\sim$8~V. Under reverse bias, the polarization-induced threshold voltage is measured at $\sim-4$~V. These resonant and threshold voltages are well explained with the polarization field which is opposite to that of the metal-polar counterpart, confirming the N-polarity of the RTDs. When the device is biased in the NDC-region, electronic oscillations are generated in the external circuit, attesting to the robustness of the resonant tunneling phenomenon. In contrast to metal-polar RTDs, N-polar structures have the emitter on the top of the resonant tunneling cavity. As a consequence, this device architecture opens up the possibility of seamlessly interfacing—via resonant tunneling injection—a wide range of exotic materials with III-nitride semiconductors, providing a route to explore new device physics.  

\end{abstract}


\maketitle
Resonant tunneling transport in III-nitride heterostructures has been under scrutiny over the last two decades~\cite{Kikuchi2001,Kikuchi2002,Golka2006,Bayram2010,sakr2012resonant,li2012repeatable,grier2015coherent}. However, only during the last four years robust quantum interference effects and room temperature negative differential conductance were reported in nitride-based double-barrier heterostructures~\cite{encomendero2016repeatable,growden2016highly,encomendero2017new,encomendero2018room,encomendero2019broken,wang2019repeatable,encomendero2020fighting,Encomendero2020Book}. Over this period, multiple advances in epitaxial growth, polar heterostructure design, device fabrication techniques, and tunneling transport theory, have been instrumental in advancing our understanding of resonant tunneling injection across polar semiconductors, leading to the realization of high-performance III-nitride resonant tunneling diodes (RTDs).

The technological importance of nitride-based resonant tunneling injection stems from the possibility of engineering the electron transport dynamics, thereby enabling the operation of ultra-high-speed electronic devices.\cite{izumi20171,kanaya2014fundamental,kanaya2015structure}
Owing to their high breakdown electric field, high longitudinal optical phonon energy, and high thermal conductivity, nitride semiconductors represent a promising platform for the development of high-speed and high-power electronic and photonic devices.\cite{cho2017single,cho2019blue}
In spite of their outstanding material properties, III-nitride semiconductors exhibit strong internal polarization fields which makes the engineering of quantum confined states, a nontrivial task.

Due to their noncentrosymmetric crystal structure, nitride heterostructures grown along the polar $c$-axis, result in a discontinuous electrical polarization $\vec{P}(z)$ which gives rise to highly localized polarization charges $q\sigma_\pi=-\Delta\vec{P}(z)\cdot\hat{z}$. Here, $\Delta\vec{P}(z)$ is the polarization discontinuity at the heterojunction interface, $q$ is the absolute value of the electron charge, and $\hat{z}$ is the unitary vector along the growth direction. The presence of these sheets of polarization charge in turn generates
strong internal spontaneous and piezoelectric polarization fields $F_\pi=q\sigma_\pi/\epsilon_s$ with magnitudes on the order of
1--10~MV/cm ($\epsilon_s$ is the dielectric constant).\cite{bernardini1997spontaneous}
Due to their strength, they modulate the spacial distribution of free carriers, determine the strength and direction of internal electric fields, and dominate the energy band profile of nitride heterostructures. This distinctive feature greatly broadens the design space of polar heterostructures via polarization engineering.\cite{jena2011polarization} 
This technique has been exploited in various photonic and electronic devices to induce 2D\cite{Khan1992,Chaudhuri2019} and 3D\cite{Jena2002,Simon2010} free carrier populations, couple electron and hole states via interband tunneling,\cite{Simon2009,krishnamoorthy2013low,Yan2015} and demonstrate a wide bandgap tunneling field-effect transistor.\cite{chaney2020gallium}

In III-nitride resonant tunneling heterostructures,
the physics of resonant injection is
greatly influenced by the interfacial polarization charges present at every heterojunction. Under equilibrium conditions, the interplay between the fixed polarization charges and mobile free carriers leads to a redistribution of electrons around the active region. This effect results in the accumulation of free electrons around the positive polarization charges $q\sigma_\pi=-\Delta\vec{P}\cdot\hat{z}>0$, on one side of the active structure. On the opposite side, the negative sign of the polarization charge (i.e. $q\sigma_\pi=-\Delta\vec{P}\cdot\hat{z}<0$) repels free carriers, inducing a depletion layer that effectively widens the adjacent tunneling barrier.
Therefore, whereas electrons on the emitter 2D electron gas (2DEG) can readily tunnel into the active region, carriers on the collector region undergo a strong attenuation by the wide depletion layer.\cite{encomendero2019broken,encomendero2020fighting} This analysis reveals that, in polar RTDs, the position of the emitter electrode, with respect to the double-barrier structure, is determined by the polarity of the crystal, which controls the sign of the polarization discontinuity $\Delta\vec{P}(z)$.

Metal-polar RTDs---grown along the [0001] direction---have the emitter buried below the double-barrier structure, thereby limiting its electrostatic control by means of surface metallic electrodes. In contrast, by flipping the polarity of the crystal, we can re-locate the emitter on the $\textit{top}$ of the resonant tunneling heterostructure. This device architecture allows un-screened control over the 2DEG population via field-effect, benefiting not only vertical but also lateral transport. 
This advantage has been recently exploited for the manufacture of highly-scaled enhancement-mode transistors with outstanding power capabilities.\cite{singisetti2013high,wong2013n,Wong2019} Using this architecture, highly-scaled transistors with 2DEG channels located at $\approx 5$~nm from the top surface, have been manufactured, attesting to the excellent electrostatic control over the 2DEG.\cite{Singisetti2012}
In addition, N-polar high electron mobility transistors (HEMTs), could be potentially readily integrated with a resonant tunneling cavity, enabling electronic gain within the terahertz band.\cite{Sensale2013,Zhao2014,Condori2016}
N-polar-based polarization engineering is also promising for the design III-nitride photocathodes\cite{Marini2018},
light-emitting diodes\cite{Simon2010,Verma2011,bharadwaj2020enhanced,turski2019polarization}
and solar cells\cite{li2011effects}
with enhanced emission, injection, and collection efficiencies, respectively.
More fundamentally, epitaxy along the
[000$\bar{1}$] direction
offers additional advantages 
stemming from the higher thermal stability of the N-polar crystal surface.\cite{VanMi2004,Togashi2009,okumura2014growth,xu2003effects}

In the case of III-nitride resonant tunneling devices, the N-polar platform allows not only an enhanced control over the source of tunneling carriers, but can also enable monolithic integration of the double-barrier structure with a variety of functional materials at the top emitter contact. From the crystal growth point of view, this is a significant advantage because highly dissimilar materials such as ferromagnets and superconductors can replace the semiconductor emitter, without compromising the structural quality of the double-barrier structure underneath. This device architecture opens up the possibility of seamlessly interfacing---via resonant tunneling injection---a wide range of exotic materials with III-nitride semiconductors, providing a route to explore new device physics.\cite{jena2019new}

In spite of their multiple advantages, N-polar resonant tunneling heterostructures have not been demonstrated so far mainly due to the lack of high-quality substrates, coupled with the difficulty of growing tunneling heterostructures on crystals containing a high density of dislocations. By virtue of advanced nitride growth technology, however, high-quality N-polar GaN substrates with low dislocation densities have recently become commercially available. 

In this paper, by taking advantage of single-crystal GaN substrates, we report the molecular-beam-epitaxy (MBE) growth, fabrication, and tunneling transport characteristics of N-polar RTDs, exhibiting robust negative differential conductance (NDC) and RF oscillations at room temperature.

\begin{figure}[t!]
\centering
\includegraphics*[width=8.5cm]{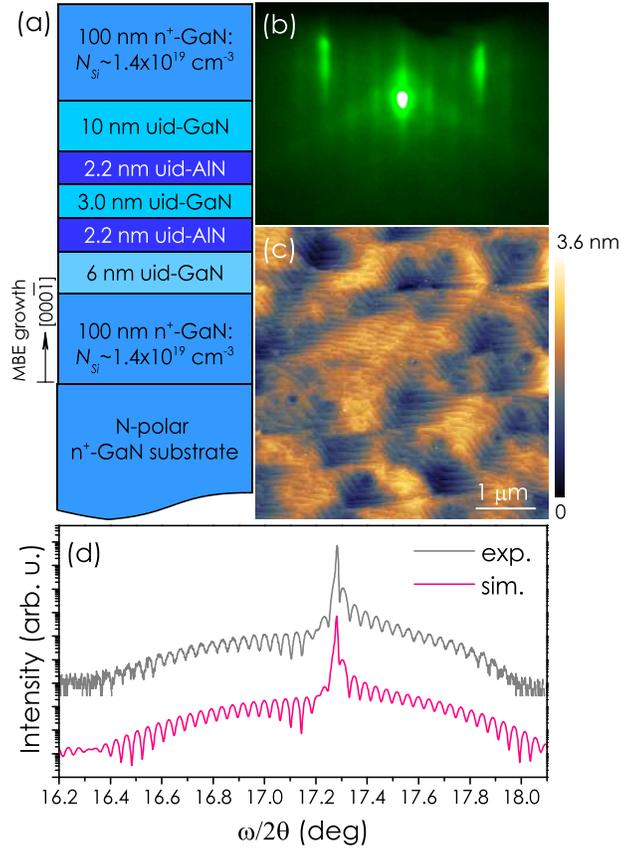} 
\caption{(a) Schematic layer structure of N-polar GaN/AlN resonant tunneling diodes. (b) The RHEED pattern, (c) $5\times5$~$\mu$m$^{2}$ AFM micrograph and (d) symmetric XRD $\omega/2\theta$ scan of the sample. The RHEED pattern has been taken below 300~\celsius along the $\langle11\bar{2}0\rangle$ azimuth after growth. The root-mean-square roughness measured by AFM on the surface in (c) is 0.50~nm.}
\label{mbe}
\end{figure}

GaN/AlN 
double-barrier heterostructures 
were grown on single-crystal N-polar GaN 
wafers---with a dislocation density of $5\times10^{4}$~cm$^{-2}$ using 
a Veeco GENxplor MBE system equipped with standard effusion cells for elemental Ga, Al and Si, and a radio-frequency plasma source for the active N species. 
The base pressure of the growth chamber was in the range of 10$^{-10}$~Torr under idle conditions, and $1.5\times10^{-5}$~Torr during growth.
The device structure consists of the following layers, starting from the nucleation surface: 100~nm GaN:Si / 6~nm GaN / 2.2~nm AlN / 3~nm GaN / 2.2~nm AlN / 10~nm GaN / 100~nm GaN:Si,
as shown in Fig.~\ref{mbe}(a).
The GaN and GaN:Si layers are grown 
under Ga-rich conditions ($\phi_{Ga}=$7.8~nm$^{-2}$~s$^{-1}$; $\phi_{N}=$4.1~nm$^{-2}$~s$^{-1}$) and the 
AlN barriers
are grown at nominally stoichiometric condition (i.e., $\phi_{Al}=\phi_{N}=$4.1~nm$^{-2}$~s$^{-1}$), under the Ga flux ($\phi_{Ga}=$7.8~nm$^{-2}$~s$^{-1}$), to ensure a metal-rich condition ($\phi_{Al}+\phi_{Ga}>\phi_{N}$); where $\phi_{Ga}$, $\phi_{Al}$ and $\phi_{N}$ are Ga, Al and active N fluxes, respectively. The entire heterostructure was grown at a constant substrate thermocouple temperature of 700~\celsius.
The excess Ga droplets after the growth were removed in HCl before \textit{ex situ} characterization and device fabrication.

Figure~\ref{mbe}(b) displays 
the reflection high-energy electron diffraction
(RHEED) pattern of the RTD sample taken at low temperature ($<300$~\celsius) after growth. 
It reveals pure reflection patterns with a well-defined specular spot and pronounced Kikuchi lines, indicating smooth surface morphology and high structural order.
More importantly, a $(3\times3)$ surface reconstruction is clearly observed, confirming that the N-polarity has been preserved up to the top layer.
Atomic force microscopy (AFM) reveals a smooth surface morphology exhibiting clear atomic steps [see Fig.~\ref{mbe}(c)]. In addition, Fig.~\ref{mbe}(d) shows the symmetric
x-ray diffraction (XRD)
scan of the
double-barrier structure. Excellent agreement between the simulated curve, based on the layer structure in Fig.~\ref{mbe}(a), and the experimental data 
is obtained. This result indicates that the GaN/AlN interfaces exhibit atomically abrupt transitions, which is critical for coherent electron injection.
\begin{figure*}[t!]
\centering
\includegraphics*[width=12cm]{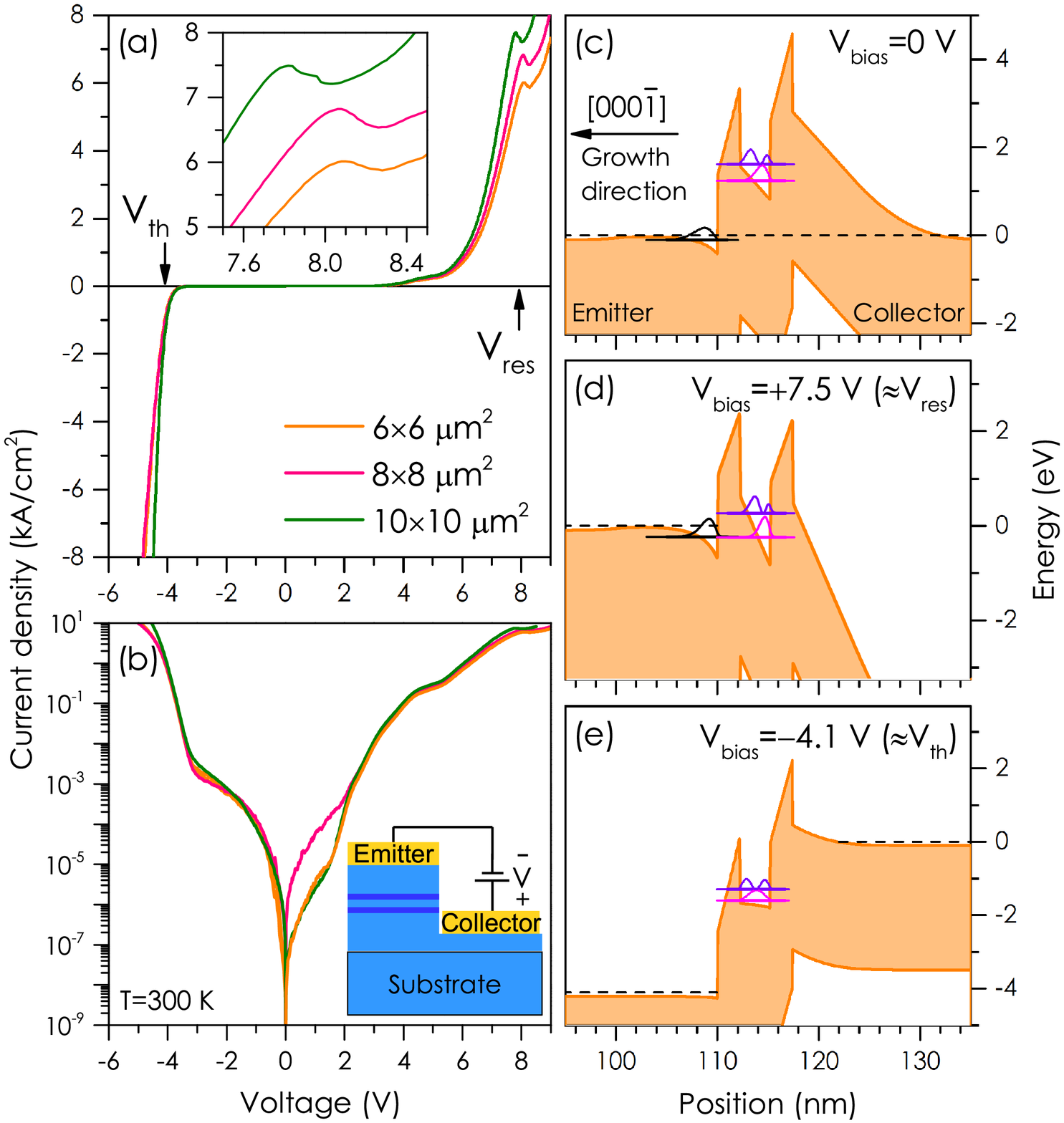} 
\caption{Current density vs voltage characteristics of N-polar GaN/AlN resonant tunneling diodes with different mesa areas, measured at room temperature in (a) linear and (b) semilogarithmic scales. Electronic transport is measured employing the test circuit depicted in the inset of panel (b). Under this configuration, forward bias corresponds to the electronic flow from the top emitter, through the double-barrier, into the collector contact at the bottom.
	The inset in (a) shows a magnification of the negative differential conductance region. Conduction-band diagrams calculated at (c) equilibrium, (d) resonant and (e) threshold voltages. The dashed lines in (c)--(e) indicate the corresponding
	emitter and collector Fermi levels. The black, magenta and violet lines designate the energy levels of the 2DEG formed at the emitter/AlN barrier, ground and first excited states in the GaN/AlN quantum well, respectively.}
\label{iv}
\end{figure*}

After growth, RTDs are fabricated by conventional contact lithography, reactive ion etching and electron-beam metal evaporation for ohmic contacts. Figures~\ref{iv}(a) and \ref{iv}(b) 
display the room-temperature current-voltage (\textit{J}-\textit{V}) characteristics of three devices with different mesa
areas, plotted in linear and logarithmic scales, respectively. The test circuit, shown schematically in the inset of Fig.~\ref{iv}(b), is set up such that the forward bias direction corresponds to electronic flow from the top emitter, through the double-barrier, into the collector contact at the bottom of the active region.
These \textit{J}-\textit{V} curves clearly resemble the electronic transport characteristics of metal-polar GaN/AlN RTDs, but with a flipped voltage bias polarity.\cite{encomendero2017new,encomendero2019broken,encomendero2020fighting}

As discussed previously, the sheets of charge $\pm q\sigma_\pi$,
located at the GaN/AlN interfaces,
induce a rearrangement of free carriers. Owing to the lack of inversion symmetry in the spatial distribution of these polarization charges, the internal electric fields inside the GaN quantum well and AlN barriers exhibit antiparallel orientations. This can be seen in the equilibrium energy band diagram of Fig.~\ref{iv}(c), calculated by solving Schr\"odinger and Poisson equations self-consistently.
More importantly, the broken symmetry in the charge distribution results in the polarization-induced broadening of the collector tunneling barrier, which strongly attenuates electron tunneling transmission. As a consequence, under low current injection 
(i.e. $|V_{bias}|<3$~V), electronic transport is supported mainly by thermally activated carriers.\cite{encomendero2017new,encomendero2020fighting}
However, as the forward bias increases (i.e. $V_{bias}>3$~V), the collector tunneling transmission grows exponentially, thereby restoring the symmetry between the emitter and collector transmission coefficients. As a consequence, constructive quantum interference within the well leads to an enhanced resonant tunneling transmission.\cite{encomendero2019broken}
When $V_{bias}\approx +4.0$~V, the emitter Fermi level gets aligned with the ground state of the well, enabling resonant tunneling injection across the active region. The detuning from this resonant condition results in the conductance modulation observed between 4 and 5~V [see Figs.~\ref{iv}(a) and (b)].
The main resonant peak occurs at $V_{res}=8.0\pm 0.2$~V, with a peak resonant tunneling current $J_{res}=6.8\pm 0.8$~kA/cm$^{2}$, measured in multiple devices across the 7~mm$\times$7~mm RTD sample [see the inset of Fig.~\ref{iv}(a)]. This result is consistent with the resonant tunneling alignment between the emitter subband and the ground state within the well, as can be seen from the band diagram shown in Fig.~\ref{iv}(d).
When the forward bias increases above the resonant tunneling voltage, the diodes exhibit a region of NDC that extends approximately over 0.2~V, resulting in a peak-to-valley-current ratio (PVCR) of $\sim1.05$ at room temperature [see the inset of Fig.~\ref{iv}(a)]. The origin of the low PVCR is attributed to the presence of leakage mechanisms across the double-barrier active region, resulting in a larger valley current and thereby degrading this important RTD metric. However as pointed out previously, the presence of these leakage paths do not prevent resonant tunneling transport within the N-polar double-barrier structure. Minimizing the magnitude of the non-resonant leakage current will require further optimization in the RTD growth conditions.

\begin{figure}[t!]
\centering
\includegraphics*[width=8.5cm]{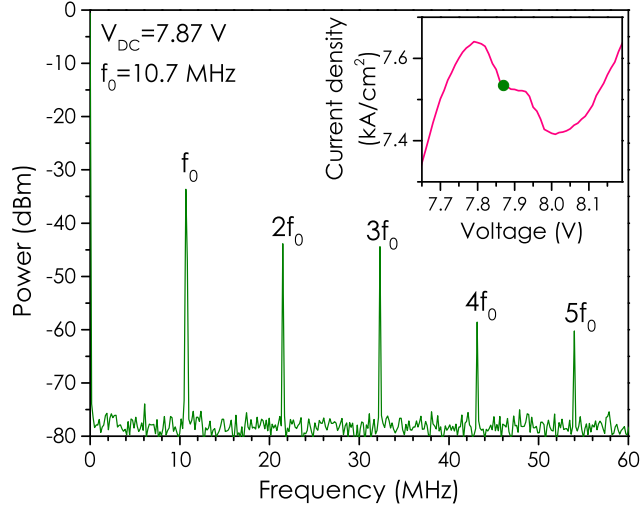} 
\caption{Power spectrum generated by the N-polar GaN/AlN resonant-tunneling-diode oscillator. The inset shows the current density vs voltage relation of the device around the negative differential conductance region. The filled circle in the inset indicates the dc bias condition used for the ac power generation.}
\label{power}
\end{figure}
Whereas forward bias injection leads to an enhanced resonant tunneling transmission, the opposite bias polarity results in an increasingly high asymmetry between the tunneling barriers.\cite{encomendero2017new,encomendero2019broken} Within this bias regime, a critical condition is achieved when electronic injection transitions from double-barrier resonant tunneling transport to single-barrier tunneling injection. This critical condition can be seen in Fig.~\ref{iv}(e) and occurs when polar RTDs are biased at the threshold voltage $V_{th}=-2t_bF_{\pi}$,
where 
$t_{b}$ is the thickness of the tunneling barriers.\cite{encomendero2017new} 
From Fig.~\ref{iv}(a), we measure the characteristic threshold voltage $V_{th}\approx-4.23$~V, using the method described in Ref.~\onlinecite{encomendero2019broken}.
Employing the barrier thickness $t_b=2.2$~nm, measured from the XRD pattern, we obtain the magnitude of the internal polarization fields along the $-c$-direction to be $F^{-c}_\pi\approx 9.6$~MV/cm. This experimental result is in reasonable agreement with previous theoretical calculations.\cite{bernardini1997spontaneous} 

To conclusively demonstrate the robustness of the resonant tunneling phenomena in our devices, we construct an oscillator circuit that exploits the room temperature NDC of the RTDs as the gain mechanism. It should be noted that the critical condition for the generation of high-speed electronic oscillations is that $G_{RTD}<-RC/L$, where $G_{RTD}$ is the RTD conductance; $R$, $C$, and $L$ are the series resistance, capacitance, and inductance of the biasing circuit, external to double-barrier structure~\cite{Hines1960}. Since the absolute value of $G_{RDT}$ is proportional to the mesa area of the device, RTDs with larger areas and thereby higher current levels are employed for the assembly of the oscillator. The transition through the critical oscillation condition can be seen in the inset of Fig.~\ref{iv}(a). The device, featuring the largest mesa area, generates oscillations in the external measurement circuit, which manifests in the chair-like shape within the NDC region. In contrast, the smaller area devices in the same figure, do not present the chair-like feature which indicates that no ac oscillations are generated. This condition can be met either by scaling the RTD area or by biasing-circuit stabilization techniques published elsewhere~\cite{Sollner1984,Shewchuk1985,Cornescu2019}.

The oscillator consists of a single N-polar RTD, with an area of $12\times12$~$\mu$m$^2$, connected to a dc voltage source and spectrum analyzer via a bias tee. When the diode is biased within the region of the NDC---shown in the inset of Fig.~\ref{power}---
self-oscillations build-up in the external circuit.
Figure~\ref{power} shows the power spectrum generated by 
resonant tunneling oscillator when
the device is biased at $V_{bias}=7.87$~V.
Owing to the 
non-linear characteristics of the differential conductance, the
output spectrum contains not only the fundamental frequency $f_{0}$,
but also multiple harmonics up to the fifth overtone.\cite{Sollner1988,Asada2001}
The output power
and frequency of the 
fundamental component,
measured at 0.43~$\mu$W and 10.7~MHz respectively, are determined by the external biasing circuit instead of the intrinsic frequency response of the RTD.\cite{encomendero2018room} Additionally, we would like to highlight that the generation of continuous and stable electronic oscillations from our devices confirm the repeatable behavior of the NDC. Therefore, under operation, the N-polar RTD-oscillator generates an ac signal which scans the NDC region at a rate of approximately $10^7$~sweeps per second, conclusively confirming the repeatability of the resonant tunneling phenomenon.


In summary, we experimentally demonstrated that resonant tunneling transport can be engineered in
GaN/AlN double-barrier heterostructures
grown along the [000$\bar{1}$] direction.
Electronic transport at room temperature, reveals 
a peak resonant tunneling current $J_{res}=6.8\pm 0.8$~kA/cm$^2$
and a resonant bias
$V_{res}=8.0\pm 0.2$~V.
When the devices are biased within the NDC region, electronic oscillations are generated in the external circuit, attesting to the robustness of the resonant tunneling phenomenon.
These results constitute the conclusive demonstration of
room-temperature resonant tunneling injection in
N-polar RTDs capable of ac power generation.
Finally, it should be noted that in contrast to metal-polar RTDs, N-polar structures have the emitter on the top of the resonant tunneling cavity. As a consequence, this device architecture opens up the possibility of seamlessly interfacing—via resonant tunneling injection—a wide range of exotic materials with III-nitride semiconductors, providing a route to explore new device physics.

The authors thank Zexuan Zhang for useful discussion. This work was supported in part by the 
AFOSR (No. FA9550-17-1-0048), NSF DMREF (No. 1534303), NSF RAISE TAQs (No. 1839196), the Semiconductor Research Corporation (SRC) Joint University Microelectronics
Program (JUMP), NSF NewLaw
(No. EFMA-1741694), and ONR (Nos. N00014-20-1-2176 and N00014-17-1-2414). This work made use of
the shared facilities that are supported through the NSF ECCS-1542081, NSF DMR-1719875, and NSF DMR-1338010.

\bibliographystyle{aipnum4-1}
\bibliography{references}
\end{document}